\documentstyle[fleqn,epsfig]{article}
\begin{document}
\title{About EAS size spectra and primary energy spectra
in the knee region}
\author{S.V.~Ter-Antonyan\footnote{e-mail: samvel@jerewan1.yerphi.am} , 
L.S.~Haroyan \\[0.5cm] 
\emph{Yerevan Physics Institute, Alikhanyan Brothers 2,
Yerevan 375036, Armenia}}
\date{}
\maketitle
\begin{abstract}
Based on the unified analyses of
KASCADE, AKENO, EAS-TOP and ANI EAS size spectra,
the approximations of
energy spectra of different primary nuclei have been found.
The calculations were carried out using the SIBYLL
and QGSJET interaction models in 0.1-100 PeV primary energy range.
The results point to existence of both rigidity-dependent steepening
energy 
spectra at $R\simeq200-400$ TV and an additional proton (neutron)
component  
with differential energy spectrum 
$(6.1\pm0.7)\cdot10^{-11}(E/E_{k})^{-1.5}$ 
(m$^2\cdot$s$\cdot$sr$\cdot$TeV)$^{-1}$  
before the knee $E_{k}=2030\pm130$ TeV
and with power index $\gamma_{2}=-3.1\pm0.05$ after the knee.
\end{abstract}
\vspace{0.5cm}

\newpage
\section{Introduction}
High statistical accuracy in modern EAS experiments in the knee region
encouraged the investigation of the fine structure of EAS size spectra. 
Although the origin of the knee is still a matter of debate, recently
the series of publications \cite{EW} appeared, where the sharpness and 
the spectral structure in the knee region were interpreted
by the contribution of heavy nuclei from a single local supernova. 
Along with this, the absolute differential EAS size spectra measured at 
different atmosphere depths and different zenith angles are not
explained yet from the point of view of a single $A-A_{Air}$ 
interaction model and a single model of primary energy spectra 
and elemental composition. Such an attempt has been made in
\cite{KHR} based on the QGS interaction model and rigidity-dependent 
steepening primary energy spectra \cite{Peters} for a description of vertical 
MSU, AKENO and Tien-Shan EAS size spectra. In the present work 
we worked out a formalism of the inverse problem solution -
reconstruction of the primary energy spectrum and elemental composition
based on the known EAS size spectra of KASCADE \cite{KAS}, AKENO \cite{AKE}, 
EAS-TOP \cite{TOP} and ANI \cite{ANI} measured at different zenith angles. 
The calculations were done in the frames of QGSJET \cite{QGS} and SIBYLL
\cite{SIB} 
interaction models. As a primary spectrum we have tested the 
modified rigidity-dependent steepening primary energy spectra and the 
hypothesis of the additional component in the knee region 
\cite{JK,JL,TG,PB2}. 
In this case the type 
of nucleus of the additional component was considering as
unknown and determining 
by the best fit of the fine structure of EAS size spectra in the knee
region.
\section{EAS inverse problem}
In general, the energy spectra ($\partial
\Im_{A}/\partial E_{0}$) of primary nuclei ($A$) 
 and detectable EAS size spectra ($\partial I
 / \partial N_{e}^{*}$) are related by the integral equation -
\begin{equation}
\frac{\partial I(E_{e},\overline{\theta},t)} {\partial N_{e}^{*}}=
\sum_{A} 
\int_{E_{min}}^{\infty}
\frac{\partial \Im_{A}} {\partial E_{0}}
W_{\theta} (E_{0},A,N_{e}^{*},\overline\theta ,t)
dE_{0}
\end{equation}
where $E_{0},A,\overline{\theta}$ are energy, nucleon number (1-59) and
average zenith angle
of primary nuclei, $E_{e}$ is an energy threshold of detected EAS
electrons, $N_{e}^*(E>E_{e})$ is the estimation value of EAS size
obtained by the electron lateral distribution function.
Here, by the EAS size ($N_{e}(E>0)$) we mean the total number of EAS
electrons
at given observation level $(t)$. The kern ($W_{\theta}$) of integral 
equation (1) is determined as 
\begin{displaymath}
W_{\theta}\equiv
\frac{1} {\Delta_{\theta}}
\int_{\theta _1}^{\theta _2}
\int_{0}^{\infty}
\frac{\partial\Omega(E_{0},A,\theta ,t)}{\partial N_{e}}
P_{\theta}
\frac{\partial \Psi(N_{e})}{\partial N_{e}^{*}}
\sin\theta d\theta dN_{e}
\end{displaymath}
where $\partial\Omega / \partial N_{e}$  is an EAS size spectrum at 
the observation level $(t)$ for given $E_{0},A,\theta$ parameters 
of a primary nucleus and depends on $A-A_{Air}$ interaction model;
$\Delta_{\theta}=\cos\theta_{1}-\cos\theta_{2}$;
\begin{displaymath}
P_{\theta}\equiv P(Ne,E_{0},A,\theta)=
\frac{1}{X\cdot Y}
\int\!\!\int D(N_{e},E_{0},A,\theta,x,y)dxdy
\end{displaymath}
is a probability to detect an EAS by scintillation array
at EAS core coordinates $|x|<X/2$, $|y|<Y/2$ and to obtain the estimations
of  
EAS parameters ($N_{e}^{*}$, $s$ - shower age, $x^{*},y^{*}$ - shower core 
location)
with given accuracies; $\partial \Psi / \partial N_{e}^{*}$ is a distribution 
of $N_{e}^{*}(N_{e},s,x,y)$ for given EAS size ($N_{e}$).\\  
One may achieve significant simplification of equation (1)    
providing the following conditions during experiments:\\
a) selection of EAS cores in a range where $P_{\theta}\equiv 1$,\\ 
b) the log-Gaussian form of the measuring error 
($\partial\Psi /\partial N_{e}^{*}$) with an average value 
$\ln(N_{e}\cdot\delta)$ and a RMSD $\sigma_{N}$, 
where $\delta$ involves all transfer factors (an energy threshold
of detected EAS electrons, $\gamma$ and $\mu$ contributions) and
slightly depends on $E_{0}$ and $A$,\\  
c) transformation (standardization) of the measured EAS size spectra to
the EAS size spectra at observation level
\begin{displaymath}
\frac{\partial I(0,\overline{\theta},t)} {\partial N_{e}}\simeq
\eta\frac{\partial I(E_{e},\overline{\theta},t)} {\partial N_{e}^{*}}, 
\end{displaymath}
where
$\eta=\delta^{(\gamma_{e}-1)}\exp\{(\gamma_{e}-1)^{2}\sigma_{N}^{2}/2\}$
and  $\gamma_{e}$ is the EAS size power index,\\
d) consideration of either all-particle primary energy spectrum 
$\partial \Im_{\Sigma}/\partial E_{0}$ with
effective nucleus $A_{eff}(E_{0})$ or energy spectra of primary nuclei
($\partial \Im _{A_{\xi}}/\partial E_{0}$)
gathered in a limited number of groups 
($\xi=1,\dots\xi_{\max}$) as unknown functions.\\
(a-d) conditions make EAS data of different experiments more 
comparable and equation (1) converts to the form 
\begin{equation}
\frac{\partial I(0,\overline{\theta},t)} {\partial N_{e}}=
\eta\int_{E_{min}}^{\infty}
\frac{\partial \Im_{\Sigma}} {\partial E_{0}}
\frac{\partial\Omega(E_{0},A_{eff}(E_{0}),\overline{\theta},t)}{\partial N_{e}}
dE_{0}
\end{equation}
or
\begin{equation}
\frac{\partial I(0,\overline{\theta},t)} {\partial N_{e}}=\eta
\sum_{\xi=1}^{\xi_{\max}}
\int_{E_{min}}^{\infty}
\frac{\partial \Im _{A_{\xi}}}{\partial E_{0}}
\frac{\partial\Omega(E_{0},\overline{A}_{\xi},\overline{\theta},t)}{\partial N_{e}}
dE_{0}
\end{equation}
However, even in this form the determination of primary energy spectra 
by measured EAS size spectra and solution 
of integral equations (2,3) in general is unsolvable problem.
At the same time, using the a priori information about energy spectra 
of primary nuclei ($\partial\Im_{A_{\xi}}/\partial E_{0}$)
and the EAS size spectra $\partial I/ \partial N_{e}^{*}\equiv
 f_{i,j}(N_{e,i}^{*},\overline{\theta}_{j},t)$ measured in $i=1,\dots m$
size
intervals and $j=1,\dots n$ zenith angular intervals, one may transform
the
inverse problem into $\chi^{2}$-minimization problem 
\begin{equation}
\min\{\chi^{2}\}\equiv
\min\Big\{
\sum_{i}^{m}\sum_{j}^{n}
\frac{(f_{i,j}-F_{i,j})^{2}}{\sigma_{f}^{2}+\sigma_{F}^{2}}
\Big\}
\end{equation} 
with unknown (free) spectral parameters.    
Here $F_{i,j}\equiv F(N_{e,i}^{*},\overline{\theta}_{j},t)$
 are the expected EAS size spectra determined at the right
hands of equations (1-3) and $\sigma_{f},\sigma_{F}$ are the
uncertainties (RMSD) of measured ($f_{i,j}$) and expected ($F_{i,j}$) 
shower size spectra.\\
One may also unify the data of different experiments applying 
minimization $\chi^2_{U}$ with re-normalized EAS size spectra  
\begin{equation}
\min\{\chi^2_{U}\}\equiv
\min\Big\{\chi^{2}\Big(
\frac{f_{i,j,k}}{\sum_{i}\sum_{j} f_{i,j,k}},
\frac{F_{i,j,k}}{\sum_{i}\sum_{j} F_{i,j,k}}
\Big)\Big\}
\end{equation}     
where index $k=1,\dots l$ determines the observation levels ($t$)
of experiments. Expression (5) offers an advantage for experiments
where the values of methodical shift ($\delta$) and measuring error
($\sigma_{N}$) are unknown or known with insufficient accuracy.\\
The energy spectra of primary nuclei are preferable to determine (a
priori) in the following generalized form
\begin{equation}
\frac{\partial \Im_{A}} {\partial E_{0}}\simeq\beta\cdot\Phi_{A}\cdot 
E_{0}^{-\gamma_{1}(A)}\cdot  
\Big(1+ \Big(\frac{E_{0}}{E_{knee}(A)}\Big)^{\epsilon}\Big)^
{(\gamma_{1}(A)-\gamma_{2})/\epsilon}
\end{equation}
Unknown (free) spectral parameters in approximation (6) 
are $\beta$, 
$E_{knee}(A)$ (so called "knee" of energy spectrum of $A$ nucleus), 
$\gamma_{1}$ and $\gamma_{2}$ (spectral asymptotic slopes before and after
knee),
$\epsilon$ (sharpness parameter of knee, $1\leq \epsilon \leq 10$). 
The values of $\Phi_{A}$ and $\gamma_{1}(A)$ parameters 
are known from approximations of balloon and satellite data \cite{PB1}
at $A\equiv 1,4,\dots 59$ and $E_{0}\simeq 1-10^{3}$ TeV. Parameter 
$\beta\simeq 1$ determines the normalization of spectra (6) in 
$10^{2}-10^{5}$ TeV energy range.\\
Thus, minimizing $\chi^2$- functions (4,5) on the basis of
measured values of $\partial I(\overline{\theta}_{i,k})/ \partial N_{ej,k}$
and corresponding expected EAS size spectra (2,3)
at given $m$ zenith angular intervals, $n$ EAS size 
intervals and $l$ experiments one may evaluate the parameters of the
primary spectrum (6). 
Evidently, the accuracies of solutions for spectral parameters strongly
depend on the number of measured intervals ($m\cdot n\cdot l$),
statistical errors and correctness of
$\partial\Omega(E_{0},A,\theta,t)/\partial
N_{e}$ determination in the framework of a given interaction model.

\section{Results}
Here, the parametric solutions of the EAS inverse problem are obtained
on the basis of KASCADE \cite{KAS} ($t=$1020 g/cm$^{2}$), AKENO \cite{AKE}
(910 g/cm$^{2}$), EAS-TOP \cite{TOP} (810 g/cm$^{2}$) and ANI \cite{ANI}
(700 g/cm$^{2}$) published EAS size spectra. 
These experiments were carried out at different observation levels and were
chosen for two reasons: satisfaction of (a-c) conditions from the section 
(2)
and a high statistical accuracy of presented data (especially KASCADE 
experiment).
Unfortunately, during the standardization of EAS size spectra (condition
(c))
the value of $\eta$ parameter is not always known with proper accuracy,
 which is the main reason of discrepancy in the results of 
different experiments. In our calculations we included $\eta$ in the list 
of unknown spectral parameters and determined by the minimization of 
functional (4). The problem does not exist if the minimization of 
re-normalized EAS size spectra is unified, since the linear parameters
($\eta_{k}\cdot\beta$) are canceled out from the functional (5).\\
The differential EAS size spectra $\partial\Omega (E_{0},A,\theta ,t)/
\partial N_{e}$ for given 
$E_{0}\equiv 0.032,0.1,\dots,100$ PeV, 
$A\equiv 1,4,12,16,28,56$,
$t\equiv 0.5,0.6,\dots,1$ Kg/cm$^{2}$,
$\cos\theta\equiv 0.8,0.9,1$
were calculated using CORSIKA562(NKG) EAS simulation code \cite{COR} at 
QGSJET \cite{QGS} and SIBYLL \cite{SIB} interaction models. 
Intermediate values are calculated using
4-dimensional log-linear interpolations. The estimations of errors of 
the expected EAS size spectra $\partial\Omega /\partial N_{e}$ 
at fixed $E_{0},A,\theta,t$ parameters did not exceed $3-5\%$.\\
The basic results of minimizations (4,5) at a given number ($\nu$) of
unknown spectral parameters and the values of $\chi^{2}/q$ 
(or $\chi_{U}^{2}/q_{u}$ for unified data),  
are presented in Tables~1-4, where $q=mn-\nu-1$
and $q_{u}=\sum_{k}(mn)_{k}-\nu-1$ are corresponding degrees of freedom. 
The upper (lower) rows of each experiment in Tables~1 and 4 
correspond to parameters obtained by the QGSJET (SIBYLL) interaction
model. 
\subsection{Test of rigidity-dependent energy spectra}
Table~1 contains the approximation values of spectral parameters at
rigidity-dependent approach \cite{Peters}
\begin{equation}
E_{knee}(A)=R\cdot Z
\end{equation}
where $Z$ is a charge of $A$ nucleus and R is a 
parameter of magnetic rigidity (or a critical (cutoff) energy 
in more modern model \cite{PB2,PB1}).
The results were obtained by minimizations (4,5) applying (6,7).
The magnitudes of $\gamma_1(A)$ for all nuclei were taken from \cite{PB1}.
The number of unknown (free) parameters is equal to $\nu=4$ and
corresponding
solutions at two interaction models are presented in Table~1.\\
\begin{table}
\begin{center}
\begin{tabular}{|r|c|c|c|c|l|}
\hline
Experiment& R [TV]   &$\gamma_{2}$  &$\epsilon$&$\eta\cdot\beta$&$\chi^{2}/q$\\
\hline
KASCADE&2390$\pm$190&3.46$\pm$0.12&2.2$\pm$0.3 & 1.05$\pm$0.08&1.3 \\ 
\cline{2-6} 
$m\cdot n=24\cdot5$&2310$\pm$220&3.45$\pm$0.12& 1.8$\pm$0.2 & 0.69$\pm$0.05& 3.0 \\ 
\hline
AKENO  &3150$\pm$120&3.50$\pm$0.14& 10$\pm$7.0 & 1.98$\pm$0.06& 2.2 \\ 
\cline{2-6} 
$m\cdot n=20\cdot3$ &2820$\pm$110 &3.50$\pm$0.31& 10$\pm$? & 1.48$\pm$0.04& 3.1 \\ 
\hline
EAS-TOP&1450$\pm$120 &3.35$\pm$0.11&2.3$\pm$0.5 & 1.43$\pm$0.03& 1.2 \\ 
\cline{2-6} 
$m\cdot n=24\cdot5$&1540$\pm$205&3.35$\pm$0.20  & 1.4$\pm$0.3 & 1.15$\pm$0.04& 0.5 \\ 
\hline
ANI&2030$\pm$245 &3.47$\pm$0.18& 2.1$\pm$0.5 & 1.07$\pm$0.02& 0.8 \\ 
\cline{2-6} 
$m\cdot n=23\cdot3$ &2230$\pm$320&3.49$\pm$0.23 & 1.9$\pm$0.4 & 0.87$\pm$0.02& 1.0 \\ 
\hline
Unified data    &2610$\pm$710&3.47$\pm$0.23 & 1.3$\pm$0.2 & - & 1.7\\
\cline{2-6} 
$\sum m\cdot n=369$&3000$\pm$650&3.49$\pm$0.30 & 1.2$\pm$0.1 & - & 2.3\\
\hline
\end{tabular}
\caption{Rigidity ($R$), slope ($\gamma_{2}$), "sharpness"  
($\epsilon$), shift ($\eta\cdot\beta$) and corresponding $\chi^2/q$ values
obtained by approximations of EAS size spectra at 
QGSJET (upper rows) and SIBYLL (lower rows) interaction models
and rigidity-dependent assumption (7).}
\end{center}
\end{table}
The stability of solutions (or the steep of $\chi^2$ global minimum)
of minimizations (4,5) is seen from obtained errors. So, it is seen that
the $\chi^2$ minimum  for AKENO data does not depend on the  
$\varepsilon$ sharpness parameter at SIBYLL and partly at QGSJET
interaction models. \\
The obtained slopes of primary energy spectra 
after knee agree with the same calculations \cite{KHR} performed 
by QGS model and exceed well known expected values ($3-3.1$) 
in the $\sim10^{17}$ eV energy range \cite{PB1}. 
Such a steep of primary spectra after the knee is a result of
$(\gamma_2,R)$ correlations in (4-7).
In case of $R\simeq2000$ TeV the $E_k(Fe)\simeq 5.2\cdot 10^4$ TeV and
the primary energy spectra in the large interval ($E_k(H)-E_k(Fe)$) at
fixed $\gamma_1(A_\xi)$ can be conformed with corresponding EAS size spectra
provided abnormally steep slope ($\gamma_2\simeq3.4-3.5$) after the
knee.\\
The results of expected EAS size
spectra in comparison with corresponding experimental data 
are shown in Fig.~1 (the thin solid lines by QGSJET model, the thin dashed
lines by SIBYLL model). 
Despite the satisfactory agreement ($\chi^2\sim1$) of EAS size spectra
with predictions of rigidity-dependent steepening spectra (6,7)
and QGSJET interaction
model the form (fine structure) of the measured EAS size spectra in the knee   
region differs from the form of corresponding expected spectra. 
It is worth mentioning that the difference is formally small and does not exceed
several percents in precise KASCADE data.\\
As a next step, we attempted to test the
rigidity-dependent approximation (7) directly.
The knee-energies $E_k(A_{\xi})$ were chosen as unknown (free) 
parameters in the approximations of EAS data (4-7), 
where $\xi=1,\dots \xi_{\max}$. 
However, the stable solutions for free parameters
were obtained only at unified minimization (5), fixed values
$\gamma_1(A_{\xi}),\gamma_2,\varepsilon$ 
and number of nuclear groups $\xi_{\max}\le 5$.
In Table~2 the values of spectral parameters $E_{knee}(A_{\xi})$ obtained 
by minimization of $\chi_{U}^2$ (5) for 5 groups 
of primary nuclei ($H$), ($He,Li$), ($Be-Na$), ($Mg-Cl$), ($Ar-Ni$) 
at given values of $\gamma_{2}=3.1$ , $\epsilon=4.0$ and $\nu=5$ are
presented. 
It is seen, that approach (7) 
is performed only for nuclei with $A>1$ and $R\simeq400$ TV
or (7) is valid for all nuclei at $R\simeq400$ TV 
but there is an additional proton flux with 
$E_{knee}^{(p)}\simeq3000$ TeV which shifts the
knee value of the total proton energy spectrum. \\
\begin{table}
\begin{center}
\begin{tabular}{|r|c|c|c|c|c|l|}
\hline
Model&$E_{k}({\emph{H}})$[TeV]&$E_{k}({\emph{He,Li}})$&$E_{k}({\emph{Be-Na}})$
&$E_{k}({\emph{Mg-Cl}})$&$E_{k}({\emph{Ar-Ni}})$&$
\chi_{U}^{2}/q_{u}$\\  
\hline
QGSJET&3070$\pm$160&790$\pm$70& 5550$\pm$60&6310$\pm$50& 9020$\pm$170&1.7\\
\hline
SIBYLL&3150$\pm$90&610$\pm$30& 6060$\pm$65 &6970$\pm$40&9780$\pm$90&2.3\\
\hline
\end{tabular}
\caption{Spectral parameters $E_{knee}(A)$ $[TeV]$ 
for different groups of primary nuclei
and different interaction models. The results obtained by unified 
analyses of EAS data at $\gamma_{2}=3.1$ and $\epsilon=4.$}
\end{center}
\end{table}
The following testing of the  rigidity-dependent
approach (7) is based on the investigation of the all-particle energy
spectrum. Toward this end the
fits of primary energy spectra $d\Im_A/dE_0$ 
for different nuclei ($A=1,\dots 59$)  known from
\cite{PB1} 
were extrapolated up to $E_{A}=10^5$ TeV energies taking into account
(6,7) at $R\simeq 600$ TeV \cite{PB2}. 
The obtained expected all-particle energy spectrum 
$dI_{\Sigma}/dE_0=\sum_{A} d\Im_A/dE_0$ was approximated by expression
similar to (6) at five free parameters 
($\Phi_\Sigma,\gamma_1,\gamma_2,E_k,\varepsilon$).
The average values of primary nuclei ($\exp(\overline{\ln A})$)
obtained from extrapolations of spectra \cite{PB1} were approximated by
step function 
\begin{equation}
A_{eff}(E_{0})=a+b\cdot\ln\big(\frac{E_{0}}{E_{k}}\big)
\end{equation}
where $b=b_{1}$ at $E_{0}<E_{k}$ and $b=b_{2}$ at $E_{0}>E_{k}$. 
The results of these approximations 
are presented in a last row of Table~3 . 
The value of sharpness parameter was equal to $\varepsilon=1\pm0.1$.\\
The first and second rows of Table~3 content the
parameters of the all-particle energy spectra 
($\partial \Im_{\Sigma}/\partial E_{0}$), 
which were obtained by minimization $\chi_{U}^2$ (expressions 2,5-7)
of unified EAS size data at $\nu=6$ and
$\varepsilon=1$. The approximation (8) has been used
for $A_{eff}(E_0)$ which is at the right hand of expression (2). 
It is necessary to note that the solutions for
$a,b_1,b_2$ parameters
can be obtained only by re-normalized EAS size
spectra (5) because of the strong
correlation between a linear parameter $\eta$ and
an effective nucleus $A_{eff}$.\\
It is seen from Table~3 that the model of
rigidity-dependent energy spectra predicts the
increase ($b_1>0$, third row of Table~3) of the mean
nucleus with energy, whereas the presented analysis of
EAS data points out a decrease of $A_{eff}$ with energy
($b_1<0$, first two rows of Table~3) in the energy range of $E_0<E_k$. 
It is obvious, that despite $\overline{A}$ could not be exactly
equal to $A_{eff}$, at least their dependence on energy
must be the same. \\
The results of recent precise experiments DICE \cite{DC} and
CASA-BLANKA \cite{CB} also point out to the decrease of $\overline{\ln A}$ 
with energy at $E_0<E_k$. 
This dependence of $A_{eff}$ and $\overline{\ln A}$ on energy might be
explained by the contribution of an additional light
component in a primary nuclei flux.\\
\begin{table}
\begin{center}
\begin{tabular}{|r|c|c|c|c|c|c|l|}
\hline
Model&$\gamma_{1}$ &$\gamma_{2}$ &$E_{k}$ [TeV]&$a$&$b_{1}$&$b_{2}$&$\chi^{2}_{U}/q_{u}$\\  
\hline
QGSJET&2.72$\pm$0.01&3.05$\pm$0.03&2180$\pm$110&1.71$\pm$0.14&-0.30$\pm$0.13&0.86$\pm$0.12&1.5\\ 
\hline
SIBYLL&2.79$\pm$0.02&3.09$\pm$0.02&3680$\pm$230&10.5$\pm$1.00&-0.28$\pm$0.20&4.24$\pm$0.61&1.4\\
\hline
R=600TV&2.66$\pm$0.01&3.09$\pm$0.01&3400$\pm$200&7.0$\pm$0.5
&0.25$\pm$0.15 &3.0$\pm$0.5  &1.2\\
\hline
\end{tabular}
\caption{Parameters of all-particle energy spectrum 
obtained by unified EAS data (the first and second rows).
The last row corresponds to results of extrapolation of 
fits \cite{PB1} taking into account the assumptions (6,7) at $R=600$ TV.}
\end{center}
\end{table}

From the above analyses follows that rigidity-dependent steepening energy 
spectra in combination with QGSJET or SIBYLL interaction model 
can not
explain the obtained results of the fine structure of EAS size spectra 
\cite{KAS,AKE,TOP,ANI} in
the knee region (Table~1 and Fig.~1), the large values of knee $E_{k}(H)$
for Hydrogen component (Table~2) and dependence of the effective nucleus 
$A_{eff}(E_{0})$ on primary energy before the knee (Table~3). 
In this connection we have carried on the search of more adequate model of
primary energy spectra and elemental composition. 
\subsection{Test of additional component}
Based on predictions \cite{JK,JL,TG,PB2} the primary energy spectra
in approximation (6) have been added by a new (polar cap \cite{PB2}) component 
$\partial \Im_{Add}/\partial E_{0}$ with power energy spectrum
\begin{equation}
\frac{\partial \Im_{Add}} {\partial E_{0}}=
\Phi^{(p)}\big(E_{k}^{(p)}\big)^{\gamma_{1}^{(p)}}
\Big(\frac{E_0}{E_{k}^{(p)}}\Big)^{-\gamma^{(p)}}
\end{equation}
where $\gamma^{(p)}=\gamma_{1}^{(p)}$ at $E_{0}<E_{k}^{(p)}$
  and $\gamma^{(p)}=\gamma_{2}$ at $E_{0}>E_{k}^{(p)}$.\\
New spectral parameters $\Phi^{(p)}, \gamma^{(p)},E_{k}^{(p)}$ 
and nucleon number $A^{(p)}$ of the additional component are considered
as unknown and determined together with parameters of
rigidity-dependent energy spectra (6,7) by minimization of $\chi^2$ and 
$\chi^2_{U}$ (4,5) at $\nu=7$. The results of expected EAS size
spectra for each experiment (KASCADE, AKENO, EAS-TOP and ANI)
taking into account contribution of additional component (9) are shown
in Fig.~1 (the thick solid lines by QGSJET model and the thick dashed 
lines by SIBYLL model). It is seen that the additional component
with high accuracy (2-5\% for KASCADE and ANI data) describes the
fine structure of EAS size spectra in the knee region. 
The values of slopes of additional component before the knee turned
out
to be $\gamma_{1}^{(p)}\simeq1.5\pm^{0.6}_{0.2}$.
The nucleon number ($A^{(p)}$) of this component with high reliability 
did not exceed of $A^{(p)}=1$ for most of experiments 
especially at QGSJET interaction model 
(except from AKENO ($A^{(p)}\simeq56$)).
The unified analyses of all experiments at QGSJET and SIBYLL interaction 
models also gave a proton or neutron 
($A^{(p)}=1$) composition of the additional component.
The values of other spectral parameters at   
$\gamma_{1}^{(p)}=1.5$ and $A^{(p)}=1$ are included in Table~4.\\
The obtained result disagrees with \cite{EW} where the alike component
consists of several heavy nuclei. However, our result is based on the high 
accuracy of the coincidence ($\sim2-3\%$ for KASCADE data)
of expected and measured EAS size spectra by both $\chi^2$ criterion and the   
overlapping of fine structure of spectra.\\
The comparison of parameters $\gamma_2$ from Tables~1,4
shows that spectral slopes after the knee from Table~4 
roughly overlap with the expected slope ($3-3.1$) well known
from $N_e\gg10^7$ EAS data.  This can be explained the fact that
the spectral brake of summary proton component
($E_k^{(p)}\simeq2\cdot10^3$ TeV) is closer to the
iron component ($E_k(Fe)\simeq10^4$ TeV) at $R\simeq400$ TV.\\
The coexistence of rigidity-dependent primary energy spectra and   
additional proton flux with spectral parameters  $E_k^{(p)}\simeq2000$
TeV, $\gamma_1^{(p)}\simeq1.5$, $\gamma_2^{(p)}=\gamma_2\simeq3.1$
explains also the shifted value of the knee $E_k(H)$ in Table~2
and the decrease of $A_{eff}(E_0)$ at $E_0<E_k$ (see section 3.1).\\
Although the additional component 
does not contribute to shower sizes below or above
the knee significantly , it is essential for the reproducing of the   
sharp knee of EAS size spectra. This idea is taken from \cite{PB2} and
is confirmed here.\\
\begin{table}
\begin{center}
\begin{tabular}{|r|c|c|c|c|c|l|}
\hline
Experiment&R [TV] &$E_{k}^{(p)}$[TeV]&$\Phi^{(p)}\cdot10^6$
&$\gamma_{2}$&$\eta\cdot\beta$ &$\chi^{2}/q$\\
\hline
KASCADE&195$\pm$30&1960$\pm$150&11.6$\pm$1.0&3.15$\pm$0.04&1.11$\pm$0.02&0.8\\
\cline{2-7} 
       &150$\pm$20&1820$\pm$85&12.8$\pm$0.6 &3.14$\pm$0.03&0.74$\pm$0.07&2.3\\ 
\hline
AKENO  &290$\pm$180&4240$\pm$370&4.04$\pm$0.10&3.25$\pm$0.05&2.60$\pm$0.07&2.0\\ 
\cline{2-7} 
       &290$\pm$210&3860$\pm$310&4.59$\pm$0.16&3.24$\pm$0.03&1.86$\pm$0.02&3.0\\ 
\hline
EAS-TOP&390$\pm$150&1960$\pm$30&6.08$\pm$0.10&3.16$\pm$0.02 & 1.43$\pm$0.01&1.1\\ 
\cline{2-7} 
       &150$\pm$95&2110$\pm$130&4.37$\pm$0.50&3.11$\pm$0.04&1.41$\pm$0.04&0.3\\ 
\hline
ANI&240$\pm$130&2060$\pm$145 &6.80$\pm$0.80&3.09$\pm$0.03&1.14$\pm$0.07& 0.5 \\ 
\cline{2-7} 
         &160$\pm$85&2050$\pm$130 &6.38$\pm$0.70&3.05$\pm$0.02&1.02$\pm$0.02& 0.8 \\ 
\hline
Unified data &390$\pm$30&2030$\pm$130&5.55$\pm$0.60 &3.09$\pm$0.01 & - & 1.6\\
\cline{2-7} 
$\sum m\cdot n=369$&322$\pm$15&2150$\pm$100&4.92$\pm$0.35 &3.08$\pm$0.01 & - & 2.2\\
\hline
\end{tabular}
\caption{Spectral parameters taking into account the   
contribution of additional proton component.}
\end{center}
\end{table}

The final all-particle energy spectrum ($\partial I/\partial E_{0}$)
obtained by unified EAS size data at QGSJET interaction model
\begin{displaymath}
\frac{\partial I} {\partial E_{0}}=\beta\Big( \sum_{A}
\frac{\partial \Im_{A}} {\partial E_{0}}+
\frac{\partial \Im_{Add}} {\partial E_{0}}\Big)
\end{displaymath}
and corresponding energy spectra ($\partial \Im_{A}/\partial E_{0}$)
of 6 nuclear groups with additional component 
($\partial\Im_{Add}/\partial E_{0}$)  
at normalization $\beta=1$ \cite{PB1} are presented in Fig.~2.
The solid (dashed) line is the all-particle energy  spectrum 
obtained by unified (only KASCADE) EAS size spectra. 
Dotted lines are the energy spectra of different primary components
obtained by unified EAS data. Symbols in Fig.~2 are the data from 
DICE \cite{DC}, CASA-BLANKA \cite{CB} arrays and reviews \cite{PB1,REV}.\\ 
The numerical values of all-particle energy spectrum (solid line in
Fig.~2), corresponding error, spectrum of additional proton component
and average nucleus versus primary energy are presented in
Table~5.
\begin{table} 
\begin{center}
\begin{tabular}{|r|c|c|c|l|}
\hline
   & $\partial I/\partial E_0$ & $\Delta(\partial I/\partial E_0$)&
$\partial \Im_{Add}/\partial E_0$  & \\
$E_0$ (TeV)&  $(m^2\cdot sr\cdot s\cdot $&$\pm(m^2\cdot sr\cdot $ & $(m^2\cdot sr\cdot
s\cdot $&$\overline{\ln A}$ \\
   &  $ \cdot TeV)^{-1}$&$\cdot s\cdot TeV)^{-1}$ &$ \cdot TeV)^{-1}$&\\   
\hline
   0.200E+03 &  0.174E-06 &  0.15E-07 &  0.196E-08 &   1.71 \\
   0.250E+03 &  0.964E-07 &  0.87E-08 &  0.140E-08 &   1.72 \\
   0.312E+03 &  0.534E-07 &  0.50E-08 &  0.100E-08 &   1.73 \\
   0.391E+03 &  0.296E-07 &  0.31E-08 &  0.719E-09 &   1.73 \\
   0.488E+03 &  0.161E-07 &  0.18E-08 &  0.514E-09 &   1.76 \\
   0.610E+03 &  0.884E-08 &  0.10E-08 &  0.368E-09 &   1.78 \\
   0.763E+03 &  0.486E-08 &  0.56E-09 &  0.263E-09 &   1.79 \\
   0.954E+03 &  0.260E-08 &  0.33E-09 &  0.188E-09 &   1.81 \\
   0.119E+04 &  0.140E-08 &  0.18E-09 &  0.135E-09 &   1.82 \\
   0.149E+04 &  0.762E-09 &  0.10E-09 &  0.965E-10 &   1.81 \\
   0.186E+04 &  0.418E-09 &  0.57E-10 &  0.681E-10 &   1.79 \\
   0.233E+04 &  0.224E-09 &  0.32E-10 &  0.395E-10 &   1.81 \\
   0.291E+04 &  0.117E-09 &  0.18E-10 &  0.199E-10 &   1.88 \\
   0.364E+04 &  0.609E-10 &  0.10E-10 &  0.100E-10 &   1.94 \\
   0.455E+04 &  0.316E-10 &  0.58E-11 &  0.502E-11 &   2.00 \\
   0.568E+04 &  0.163E-10 &  0.33E-11 &  0.252E-11 &   2.05 \\
   0.711E+04 &  0.843E-11 &  0.18E-11 &  0.126E-11 &   2.11 \\
   0.888E+04 &  0.435E-11 &  0.10E-11 &  0.634E-12 &   2.16 \\
   0.111E+05 &  0.222E-11 &  0.57E-12 &  0.318E-12 &   2.19 \\
   0.217E+05 &  0.280E-12 &  0.77E-13 &  0.402E-13 &   2.19 \\
   0.827E+05 &  0.448E-14 &  0.14E-14 &  0.642E-15 &   2.19 \\
\hline
\end{tabular}
\caption{Expected 
all-particle flux, error, flux of additional component
and average nucleus at different primary energies.}
\end{center}
\end{table}
\section*{Conclusion}
High statistical accuracy of experiments KASCADE, EAS-TOP, AKENO and
ANI allowed to obtain approximations
of primary energy spectra and elemental composition 
with accuracy $\sim15\%$ in the knee region. Along with this, 
KASCADE and ANI EAS size spectra at $\theta < 37^0$ are described
with the accuracy of $\sim2-5\%$ in a whole measurement interval.\\
Obtained results
show the evidence of QGSJET interaction model at least in $10^5-10^7$ TeV
energy range, rigidity-dependent steepening primary energy spectra at 
$R\simeq200-400$ TV and existence of the additional proton (or neutron) 
component 
with spectral power index $\gamma^{(p)}_{1}\simeq 1.5\pm^{0.6}_{0.2}$ 
before the knee
$E_{k}^{(p)}\simeq 2030$ TeV. The contribution of the additional 
proton (neutron) 
component in all-particle energy spectrum turned out to be 
$20\pm5\%$ at primary energy $E_{0}=E_{k}^{(p)}$. 
\section*{Acknowledgements}
We thank Peter Biermann for extensive discussions and 
Heinigerd Rebel, Johannes Knapp and Dieter Heck for providing the CORSIKA code.

\newpage
\begin{figure}[htb]
\begin{center}
\mbox{\epsfig{file=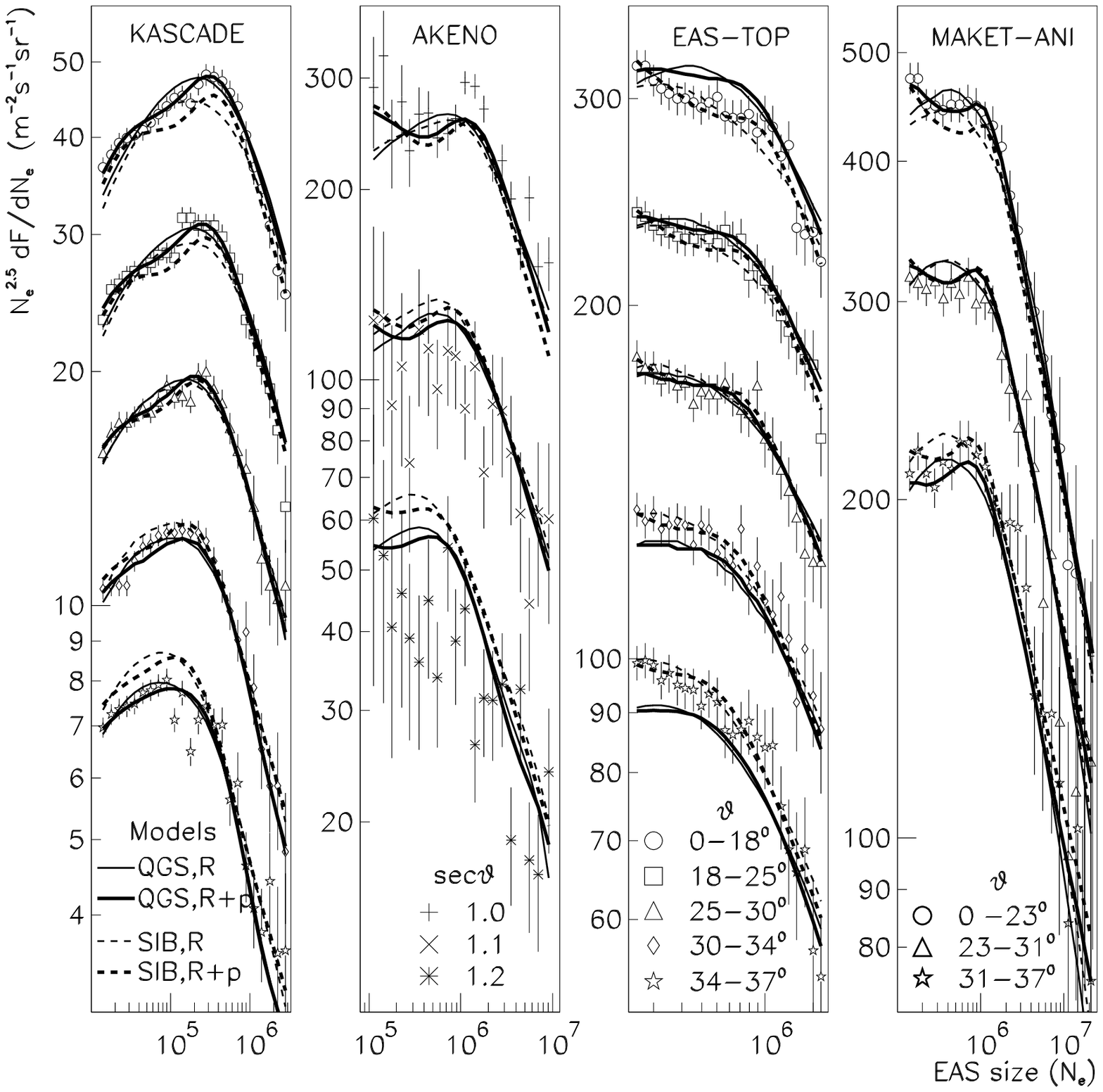,width=14cm,height=15cm}}
\end{center}
\vspace{-1cm}
\caption{KASCADE, AKENO, EAS-TOP and ANI EAS size spectra 
(symbols). The thin (thick) lines correspond to predictions
via rigidity-dependent steepening primary spectra (with the additional
proton component). The solid and dashed lines correspond to
the QGSJET and SIBYLL interaction models respectively.}
\end{figure}

\newpage
\begin{figure}[htb]
\begin{center}
\mbox{\epsfig{file=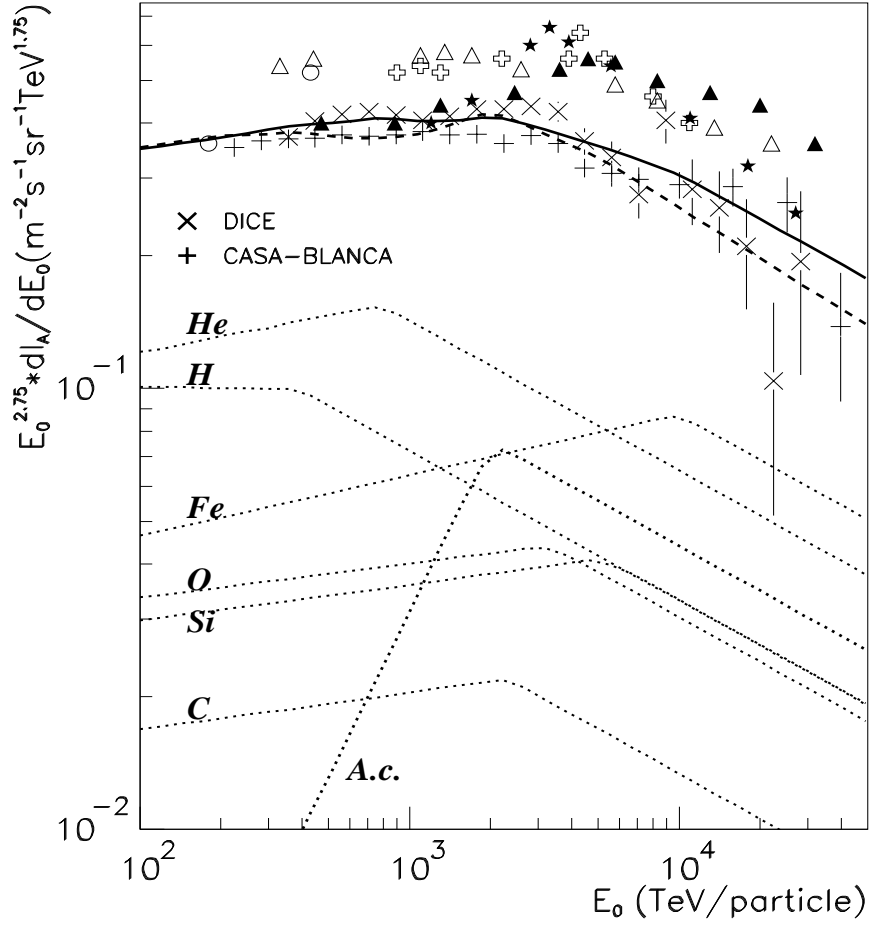,width=14cm,height=15cm}}
\end{center}
\vspace{-1cm}
\caption{Primary energy spectra and elemental composition.
The solid (dashed) line is the all-particle energy  spectrum obtained by
unified (only KASCADE) EAS data. The dotted lines are the energy spectra
of different nuclear groups. The A.c.~dotted line is the energy spectrum
of additional proton component. The symbols are the data from \cite{DC,CB}
and reviews \cite{PB1,REV}.}
\end{figure}

\end{document}